\documentclass[10pt,a4paper]{article}

\usepackage[T1]{fontenc}
\usepackage[utf8]{inputenc}
\usepackage[natbibapa]{apacite}
\usepackage[english,french]{babel}

\usepackage[babel,french=guillemets]{csquotes}
  \MakeOuterQuote{"}
\usepackage{CormorantGaramond}
\let\oldnormalfont\normalfont
\def\normalfont{\oldnormalfont\mdseries}
\usepackage[scaled=.85]{inconsolata}
\usepackage[light,tabular,lining,scale=.76]{montserrat}

\usepackage{graphicx}
\usepackage[allcolors=black,colorlinks=true,breaklinks=true,unicode]{hyperref}
\usepackage{url}
 \PassOptionsToPackage{hyphens}{url}
\usepackage{enumitem}
\usepackage{booktabs}
\usepackage{xltabular}
\usepackage{array,multirow,makecell}
 \setcellgapes{1pt}
 \makegapedcells
 \newcolumntype{R}[1]{>{\raggedleft\arraybackslash }b{#1}}
 \newcolumntype{L}[1]{>{\raggedright\arraybackslash }b{#1}}
 \newcolumntype{C}[1]{>{\centering\arraybackslash }b{#1}}
\usepackage{xltabular}

\usepackage[all]{nowidow}

\usepackage{textcomp}

\makeatletter
\newskip\@bigflushglue \@bigflushglue = -100pt plus 1fil

\def\bigcentering{\let\\\@centercr\rightskip\@bigflushglue%
\leftskip\@bigflushglue
\parindent\z@\parfillskip\z@skip}

\makeatother

\title{Réception d’une recommandation algorithmique basée sur les dispositifs de réalisation des films documentaires\thanks{\textsc{Gantier}, Samuel, \textsc{Givois}, Ève, \textsc{Jacquemin}, Bernard et \textsc{Atbane-El Houadi}, Bouchra, 2023. Réception d’une recommandation algorithmique basée sur les dispositifs de réalisation des films documentaires. \emph{Revue française des sciences de l’information et de la communication} [en ligne]. 2023. Vol.~26. [Consulté le 5 juin 2023]. DOI \href{https://doi.org/10.4000/rfsic.14204}{10.4000/rfsic.14204}.}}

\usepackage{authblk}

\author[1]{Samuel Gantier}
\author[1]{Ève Givois}
\author[2]{Bernard Jacquemin}
\author[1]{Bouchra Atbane-El Houadi}

\affil[1]{Université Polytechnique Hauts-de-France, LARSH, F-59300 Valenciennes, France.\protect\\ \{Prenom.Nom\}@uphf.fr}
\affil[2]{Univ. Lille, ULR 4073 \textendash{} GERiiCO, F-59000 Lille, France.\protect\\ \{Prenom.Nom\}@univ-lille.fr}

\date{\vspace{-3ex}}

\begin{document}

\maketitle

\begin{abstract}
\noindent Cet article analyse la réception d'une recommandation algorithmique inédite de films documentaires auprès d'un panel d'abonnés cinéphiles de la plateforme Tënk. Afin de proposer une alternative aux recommandations fondées sur une classification thématique, le réalisateur ou la période de production, un jeu de métadonnées a été élaboré dans le cadre de cette expérimentation afin de caractériser la grande variété des dispositifs de réalisation documentaires. L'enjeu est d'interroger les différentes manières dont les abonnés cinéphiles de la plateforme s'approprient une recommandation personnalisée de 4 documentaires dont les dispositifs de réalisation sont proches ou similaires. Pour conclure, les apports et limites de cette preuve de concept sont discutés afin d'ébaucher des pistes de réflexion pour améliorer la médiation instrumentée du cinéma documentaire.\\[1ex] 
\textbf{Mots-clefs:} analyse de la réception, algorithme de recommandation, plateforme de vidéo à la demande (VàD), cinéma documentaire, documentaire de création, médiation, dispositif de réalisation, cinéphilie.

\selectlanguage{english}
\begin{center}\textbf{Abstract}\end{center}
 This article analyzes the reception of a novel algorithmic recommendation of documentary films by a panel of moviegoers of the Tënk platform. In order to propose an alternative to recommendations based on a thematic classification, the director or the production period, a set of metadata has been elaborated within the framework of this experimentation in order to characterize the great variety of “documentary filmmaking dispositifs” . The goal is to investigate the different ways in which the platform's film lovers appropriate a personalized recommendation of 4 documentaries with similar or similar filmmaking dispositifs. To conclude, the contributions and limits of this proof of concept are discussed in order to sketch out avenues of reflection for improving the instrumented mediation of documentary films. \\[1ex]
\selectlanguage{french}
\end{abstract}

\section{Introduction}

Le développement massif des plateformes de vidéo à la demande lors de la dernière décennie conduit à une surabondance d'offres légales de films sur Internet. Que ce soit sur des plateformes généralistes (Netflix, Prime Vidéo, MyCanal, UniversCiné, etc.) ou spécialisées (Mubi, Dafilms, Les Yeux Doc, Spicee, Tënk, etc.), le public qui apprécie le cinéma documentaire peut consulter des catalogues composés de plusieurs milliers de titres. Or, contrairement au cinéma de fiction, pour faire face à cet hyperchoix, le spectateur ne peut pas s'appuyer sur des repères socio-sémiotiques tels que la notoriété du réalisateur\footnote{Il est bien entendu possible de suivre la carrière de quelques cinéastes confirmés mais la grande majorité de la création documentaire contemporaine est composée d'œuvres de réalisateurs peu médiatisés hors des festivals spécialisés.}, le casting des comédiens, ou la classification par genres cinématographiques \citep{delaporte_mediation_2019}.

Dans ce contexte, le programme de recherche AlgoDoc\footnote{Le programme AlgoDoc (Algorithme de recommandation de films Documentaires) a été financé par la région Hauts-de-France dans le cadre du dispositif Start Airr avec le soutien de la Meshs Hauts-de-France. Il s'appuie sur un consortium recherche-industrie composé des laboratoires LARSH de l'Université Polytechnique Hauts-de-France; du laboratoire GERiiCO de l'Université de Lille; de la société Spideo qui met à disposition son moteur algorithmique; et de la plateforme Tënk qui propose un catalogue de 2 200 films et une base de données de 20 000 abonnés.} a pour objectif de concevoir une preuve de concept qui vise à montrer qu'il est possible de recommander autrement le cinéma documentaire. Afin de proposer une alternative aux recommandations algorithmiques fondées principalement sur une classification thématique, le réalisateur ou la période de production, un jeu de métadonnées inédit a été élaboré pour caractériser les multiples agencements possibles du dispositif de réalisation documentaire. Cet article présente l'analyse de la réception de cette preuve de concept avec les utilisateurs finaux de la plateforme Tënk\footnote{Tënk diffuse depuis 2016 des documentaires d'auteurs marginalisés par la production télévisée fortement standardisée ces vingt dernières années \citep{lesaunier_conditions_2021}(Lesaunier, 2021).}. Notre problématique consiste à interroger les différentes manières dont un panel d'abonnés cinéphiles de la plateforme s'approprie\footnote{La notion d'appropriation est considérée ici comme une pratique sociale engageant \enquote{des processus cognitifs, subjectifs et culturels qui permettent de rendre “sien” un objet et met en jeu des phénomènes d'identité} \citep{jouet_usage_2017}.} les recommandations algorithmiques basées sur un dispositif de réalisation similaire. Les recommandations font-elles sens pour ce public initié au documentaire de création? Les liens de proximité entre les films sont-ils compris et jugés cohérents pour ce groupe d'abonnés cinéphiles?

Pour répondre à ces questionnements, notre démarche s'inscrit dans le sillage d'une \enquote{recherche-projet en design} \citep{findeli_rechercheprojet_2015}. Il s'agit d'adopter une méthodologie de conception \emph{exploratoire, itérative et réflexive} focalisée sur la résolution de problème \citep{gentes_indiscipline_2018,zacklad_design_2017}. Suivant ce raisonnement, notre approche interdisciplinaire en sciences de l'information et de la communication convoque trois courants de recherche. Premièrement, des travaux qui théorisent le cinéma documentaire sous un prisme sémio-pragmatique et communicationnel \citep{nichols_introduction_2010,niney_documentaire_2009,caillet_dispositifs_2014,odin_espaces_2011,lioult_enseigne_2019}. Deuxièmement, des recherches qui interrogent l'évolution historique des pratiques cinéphiles sur Internet \citep{jullier_cinephiles_2010,allard_cinephiles_2000,gimello-mesplomb_cinephilie_2015,taillibert_video_2020}. Et troisièmement, des études qui analysent les modes de recommandations algorithmiques mis en œuvre par les acteurs des industries culturelles pour valoriser leurs catalogues \citep{menard_systemes_2014,cardon_quoi_2015,vayre_histoire_2017,delcroix_donnees_2019,drumond_production_2018,beuscart_algorithmes_2019,farchy_culture_2020}.

Dans un premier temps, l'article expose la méthodologie mise en œuvre pour concevoir cette recommandation algorithmique. Dans un deuxième temps, l'analyse focalise sur la manière dont la proposition de films est collectivement interprétée par un panel de 11 abonnés cinéphiles de Tënk. Dans un troisième temps, les dynamiques d'appropriation individuelles sont explicitées afin de dégager différentes postures interprétatives. Pour conclure, les apports et limites de cette recommandation sont discutés afin d'ébaucher des pistes de réflexion sur la médiation algorithmique du cinéma documentaire.

\section{Présentation de la méthodologie expérimentale}

\subsection{Processus de conception de la preuve de concept}

La méthodologie adoptée pour concevoir cette recommandation algorithmique basée sur le dispositif de réalisation se décompose en quatre grandes étapes.

\begin{enumerate}
 \item En amont de l'expérimentation, une analyse des besoins des abonnés de Tënk a permis de construire une série personæ\footnote{Les personæ se définissent comme \enquote{des archétypes d'utilisateurs créés à partir de données réelles recueillies pendant la phase d'exploration. Ils sont utilisés dans le processus de conception pour représenter et décrire les buts, besoins, freins et caractéristiques de différents groupes d'utilisateurs.} \citep[p.~194]{lallemand_methodes_2018}. Se reporter à \citet{gantier_construction_2020} pour la présentation détaillée de la construction des personae de Tënk.}. Ce diagnostic des usages de la plateforme conduit à l'idée que les utilisateurs cinéphiles de Tënk s'intéressent davantage au dispositif de réalisation (les partis pris de mise en scène) qu'à la thématique abordée par les films (leur sujet). Ce constat constitue le postulat initial de cette étude.
 \item Le concept de \enquote{dispositif de réalisation} a ensuite été formalisé pour construire une nouvelle catégorisation du cinéma documentaire. Ce concept se définit comme un agencement complexe de différents éléments qui conditionnent la praxis du cinéaste. Le dispositif de réalisation s'avère particulièrement heuristique pour caractériser la pratique documentaire, c'est-à-dire \enquote{la stratégie de tournage, l'agencement, ou encore le protocole du film, qui structure sa mise en scène et qui concourt à la production d'un certain effet} \citep[p.~15]{caillet_dispositifs_2014}. Afin de catégoriser la grande variété des dispositifs de réalisation, un modèle théorique a permis d'identifier les multiples interactions possibles entre 6 entités: la personne filmant, la personne filmée, la situation filmée, les matériaux filmiques, le texte filmique et le public\footnote{La construction de ce modèle théorique est discutée dans une publication antérieure \citep[à paraître]{gantier_cartographier_2023}.}.
 \item Sur la base de ce modèle théorique, un thésaurus\footnote{Michèle \citet{hudon_thesaurus_2012} définit un thésaurus comme un langage documentaire dont le vocabulaire contrôlé est constitué d'un ensemble de concepts permettant de caractériser un domaine de connaissances. Se reporter à \citet{givois_elaboration_2021} pour la présentation méthodologique de la construction du thésaurus. Celui-ci est disponible en accès libre sur la plateforme OpenTheso: \url{https://opentheso.huma-num.fr/opentheso/?idt=th272} (consulté le 07/04/2023).} a été structuré afin de pouvoir indexer les différentes caractéristiques du dispositif de réalisation documentaire. Un corpus de 331 films documentaires (représentatifs du catalogue de Tënk) a ensuite été indexé par une documentaliste audiovisuelle pendant 12 mois. Chaque film a fait l'objet d'une indexation détaillée, au moyen de dix descripteurs, validés par un panel d'experts du domaine\footnote{L'ensemble du corpus indexé est détaillé dans le rapport technique de programme AlgoDoc \citep{gantier_rapport_2022}.}. En résumé, notre approche algorithmique est centrée sur la description sémantique des différents dispositifs de réalisation et se distingue totalement des démarches focalisées sur les traces de navigation des utilisateurs \citep{kembellec_moteurs_2014}.
 \item Le jeu de métadonnées a été ingéré dans l'algorithme fourni par la société Spideo de manière à calculer la distance vectorielle entre chaque film\footnote{Voir \citet{gantier_recommander_2020} pour les choix des différentes variables et pondérations adoptées dans la computation algorithmique.}. Pour conclure, l'objectif de cette preuve de concept est de guider vers de nouvelles catégories d'usage, afin d'introduire de nouvelles fonctionnalités ou variables à des systèmes algorithmiques préexistants \citep{roth_ecriture_2022}.
\end{enumerate}

\subsection{Analyse de la réception avec les utilisateurs finaux}

En 2022, la direction de Tënk revendique environ 20 000 utilisateurs dont le cœur de cible est constitué d'une communauté cinéphile qui partage un goût commun pour le documentaire d'auteur. L'hypothèse défendue dans cette étude est que ce public spécialisé constitue une communauté interprétative\footnote{ Pour l'évolution du concept de communauté interprétative dont la genèse revient à Stanley Fish se reporter à \citet{magdelaine-andrianjafitrimo_communaute_2015}.}. Selon Jean-Pierre \citeauthor{esquenazi_sociologie_2007} \enquote{participer à une communauté d'interprétation signifie maîtriser différents ensembles de catégories permettant de classer et d'appréhender un type d'œuvre donné. Les membres d'une même communauté d'interprétation disposent d'une même perspective sur l'œuvre, ce qui ne garantit pas qu'ils la comprendront de la même façon mais qu'ils invoqueront des critères analogues pour justifier le sens qu'ils lui attribuent.} \citeyearpar[p.~77]{esquenazi_sociologie_2007}. Afin de construire un échantillon représentatif de cette population, 282 abonnés ont été contactés par e-mail avec le critère commun que chaque participant ait visionné au minimum 40 documentaires sur Tënk. En février 2022, 11 entretiens directifs (5 hommes et 6 femmes) d'une durée de 90 minutes ont été menés et enregistrés en visioconférence.

{\small
 \begin{xltabular}[!ht]{\textwidth}{p{8ex}p{9ex}p{4ex}XX*{3}{p{9ex}}}
 \caption{Caractéristiques du panel d'abonnés cinéphiles de Tënk.}\label{tab:panel}\\
  \toprule
  \multicolumn{1}{C{8ex}}{N° abonné} & \multicolumn{1}{C{9ex}}{Prénom} & \multicolumn{1}{C{4ex}}{Âge} & Profession & \multicolumn{1}{C{9ex}}{Date d'abonnement} & \multicolumn{1}{C{9ex}}{Films vus du corpus} & \multicolumn{1}{C{9ex}}{Films vus sur Tënk}\\
  \midrule
  \endfirsthead
  \toprule
  \multicolumn{1}{C{8ex}}{N° abonné} & \multicolumn{1}{C{9ex}}{Prénom\footnote{Le panel a été anonymisé avec des prénoms de substitution.}} & \multicolumn{1}{C{4ex}}{Âge} & Profession & \multicolumn{1}{C{9ex}}{Date d'abonnement} & \multicolumn{1}{C{9ex}}{Films vus du corpus\footnote{Données déclaratives recueillies par questionnaire lors du recrutement du panel.}} & \multicolumn{1}{C{9ex}}{Films vus sur Tënk\footnote{Données extraites du CMS Kinow qui considère qu'un film a été vu lorsque 50\,\% de sa durée a été lue.}}\\
  \midrule
  \endhead
  A1 & David & 58 & Chercheur en mathématiques & 30 avril 2020 & 331 & 269\\
  A2 & Aline & 42 & Éducatrice spécialisée & 10 février 2020 & 128 & 174\\
  A3 & Hélène & 63 & Oléologue & 7 janvier 2019 & 51 & 43\\
  A4 & Joséphine & 33 & Étudiante en cinéma et programmatrice en festival & 19 juin 2017 & 86 & 270\\
  A5 & Marianne & 68 & Institutrice et éducatrice retraitée & 7 juillet 2017 & 63 & 260\\
  A6 & Nadia & 28 & Comédienne de théâtre & 7 janvier 2020 & 47 & 86\\
  A7 & Adeline & 39 & Productrice d'olives et réalisatrice & 11 décembre 2017 & 61 & 286\\
  A8 & Alain & 83 & Directeur retraité de centre médico-psycho-pédagogique & 1er novembre 2018 & 91 & 156\\
  A9 & Vincent & 62 & Comédien et metteur en scène de spectacle vivant & 2 janvier 2018 & 74 & 249\\
  A10 & Éric & 44 & Professeur-documentaliste & 9 novembre 2017 & 31 & /\footnote{Donnée neutralisée car l'abonné bénéficie d'un abonnement collectif.}\\
  A11 & Benjamin & 26 & Assistant-monteur & 23 février 2019 & 27 & 102\\
  \bottomrule
  \end{xltabular}
}

L'ensemble du panel partage une fréquence de visionnage importante sur Tënk. La valeur moyenne de consultation (pendant la durée variable de leur abonnement) est de 189 documentaires et la valeur médiane de 211 films. Par ailleurs, 6 abonnés sur 11 témoignent d'une fréquentation de festivals spécialisés tels que les États généraux du film documentaire (Lussas), le Mois du documentaire (France) ou encore le Cinéma du réel (Paris). Cette pratique culturelle favorise l'acquisition d'un vocabulaire idiomatique pour parler des différentes démarches cinématographiques qui composent le champ de la création documentaire. 5 abonnés sur 11 ont également participé à la fabrication d'un film ou à la construction d'une programmation dans le cadre d'une projection publique. Il s'agit par conséquent d'un public d'initiés dotés d'une littératie médiatique importante et pour certains d'un savoir-faire de praticien.

\subsection{Simulation d'un scénario d'usage}

Le protocole expérimental propose de simuler une séquence de visionnage de films documentaires diffusés par Tënk. Le point de départ consiste à demander à l'abonné de choisir un film qu'il a vu et particulièrement apprécié pour la qualité de son dispositif de réalisation. Sur la base de ce film de référence (\emph{input}), les quatre documentaires du corpus dont le dispositif est le plus proche sont recommandés à l'abonné (\emph{output}). À ces quatre œuvres, un film \enquote{contrôle}, qui ne possède aucun lien avec le film de référence, est ajouté manuellement par l'équipe de chercheurs. Chaque membre du panel dispose ainsi d'une liste de 5 documentaires classés par ordre alphabétique avec un résumé et un lien Viméo pour visionner les documentaires à l'endroit de son choix pendant un délai d'une semaine.\footnotetext{L'ensemble des films sont documentés dans la base de données du portail film-documentaire.fr.}

{\small
\begin{xltabular}[ht]{\textwidth}{p{.12\textwidth}XXX}
\caption[Recommandation personnalisée fabriquée pour l'expérimentation.]{Recommandation personnalisée fabriquée pour l'expérimentation\protect\footnotemark.\label{tab:recommandation}}\\
 \toprule
 Abonné & Film choisi  par l'abonné (\emph{Input}) & Films recommandés par l'algorithme (\emph{Output}) & Film introduit manuellement (contrôle)\\
 \midrule
\endfirsthead
 \toprule
 Abonné & Film choisi  par l'abonné (\emph{Input}) & Films recommandés par l'algorithme (\emph{Output}) & Film introduit manuellement (contrôle)\\
 \midrule
\endhead
  \begin{description}[leftmargin=*]
   \item[\hspace{-2ex}] David
 \end{description} &
  \begin{itemize}[leftmargin=*]
   \item \emph{Chris the Swiss} (Kofmel, 2018)
  \end{itemize} &
  \begin{itemize}[leftmargin=*]
   \item \emph{Carré 35} (Caravaca, 2017)
   \item \emph{Countdown} (Stonys, 2004)
   \item \emph{Klara Heydebreck\dots{}} (Fechner, 1969)
   \item \emph{Maguy Marin\dots{}} (Mambouch, 2018)
  \end{itemize} & 
  \begin{itemize}[leftmargin=*]
   \item \emph{Terrestres} (Rajotte, 2020)
  \end{itemize} \\
 \midrule
 \begin{description}[leftmargin=*]
   \item[\hspace{-2ex}] Aline
 \end{description} &
 \begin{itemize}[leftmargin=*]
  \item \emph{Dernières nouvelles\dots{}} (Bertuccelli, 2016)
 \end{itemize} &
 \begin{itemize}[leftmargin=*]
  \item \emph{Avec un intérêt\dots{}} (Reidemeister, 1983)
  \item \emph{C'est pour de faux} (Maire, 2019)
  \item \emph{Le Plein pays} (Boutet, 2009)
  \item \emph{Scheme Birds} (Hallin, Fiske, 2019)
 \end{itemize} &
 \begin{itemize}[leftmargin=*]
  \item \emph{Bab Sebta} (Maroufi 2019)
 \end{itemize} \\
 \midrule
 \begin{description}[leftmargin=*]
   \item[\hspace{-2ex}] Hélène
 \end{description} &
 \begin{itemize}[leftmargin=*]
  \item \emph{StepAcross\dots{}} (Humbert et Penzel, 1990)
 \end{itemize} &
 \begin{itemize}[leftmargin=*]
  \item \emph{Appel à l'anxiété\dots{}} (Klodawsky, 2013)
  \item \emph{Et la vie} (Gheerbrant, 1990)
  \item \emph{Fremd} (Fassbender, 2011)
  \item \emph{Gigi\dots{}} (Dervaux et Abdellaoui, 1996)
 \end{itemize} &
 \begin{itemize}[leftmargin=*]
  \item \emph{Our city} (Tarantino 2015)
 \end{itemize} \\
 \midrule
 \begin{description}[leftmargin=*]
   \item[\hspace{-2ex}] Joséphine
 \end{description} &
 \begin{itemize}[leftmargin=*]
  \item \emph{La Ronde} (Perrin, 2018)
 \end{itemize} &
 \begin{itemize}[leftmargin=*]
  \item \emph{Nuit sur Kepler\dots{}} (Voit, 2019)
  \item \emph{Libre} (Toesca, 2018)
  \item \emph{Pedra\dots{}} (Parente etVarenne, 2019)
  \item \emph{La Ravine} (Verneret, 2018)
 \end{itemize} &
 \begin{itemize}[leftmargin=*]
  \item \emph{Voisins} (McLaren, 1952)
 \end{itemize} \\
 \midrule
 \begin{description}[leftmargin=*]
   \item[\hspace{-2ex}] Marianne
 \end{description} &
 \begin{itemize}[leftmargin=*]
  \item \emph{Terrestres} (Rajotte, 2020)
 \end{itemize} &
 \begin{itemize}[leftmargin=*]
  \item \emph{Il pianeta azzurro} (Piavoli, 1982)
  \item \emph{Les Hommes} (Michel, 2006)
  \item \emph{Lightning Dance} (Bengolea, 2018)
  \item \emph{Normal} (Tulli, 2019)
 \end{itemize} &
 \begin{itemize}[leftmargin=*]
  \item \emph{Voisins} (McLaren, 1952)
 \end{itemize} \\
 \midrule
 \begin{description}[leftmargin=*]
   \item[\hspace{-2ex}] Nadia
 \end{description} &
 \begin{itemize}[leftmargin=*]
  \item \emph{Lift} (Isaacs, 2001)
 \end{itemize} &
 \begin{itemize}[leftmargin=*]
  \item \emph{Secteur 545} (Creton, 2004)
  \item \emph{Oumoun} (Ghammam, 2017)
  \item \emph{Le Village} (Simon, 2019)
  \item \emph{Les Glaneurs\dots{}} (Varda, 2002)
 \end{itemize} &
 \begin{itemize}[leftmargin=*]
  \item \emph{The Dam} (Koniarz, 2018)
 \end{itemize} \\
 \midrule
 \begin{description}[leftmargin=*]
   \item[\hspace{-2ex}] Adeline
 \end{description} &
 \begin{itemize}[leftmargin=*]
  \item \emph{Les Arbres\dots{}} (Colombo et De Luca, 2015)
 \end{itemize} &
 \begin{itemize}[leftmargin=*]
  \item \emph{Wild relatives} (Manna, 2017, 65 min)
  \item \emph{L'Esprit\dots{}} (Manchematin et Steyer, 2018)
  \item \emph{Ur-Musig} (Schläpfer, 1993)
  \item \emph{Les Poussières} (Franju, 1954)
 \end{itemize} &
 \begin{itemize}[leftmargin=*]
  \item \emph{Machini} (Tétshim, 2019)
 \end{itemize} \\
 \midrule
 \begin{description}[leftmargin=*]
   \item[\hspace{-2ex}] Alain
 \end{description} &
 \begin{itemize}[leftmargin=*]
  \item \emph{Le Village, ép. 1} (Simon, 2019)
 \end{itemize} &
 \begin{itemize}[leftmargin=*]
  \item \emph{L'Assemblée} (Otero, 2017)
  \item \emph{Cette télévision\dots{}} (Otero, 1997)
  \item \emph{Absolut Warhola} (Mucha, 2001)
  \item \emph{Wolves at the border} (Páv, 2020)
 \end{itemize} &
 \begin{itemize}[leftmargin=*]
  \item \emph{Allegro\dots{}} (Froment, 2017)
 \end{itemize} \\
 \midrule
 \begin{description}[leftmargin=*]
   \item[\hspace{-2ex}] Vincent
 \end{description} &
 \begin{itemize}[leftmargin=*]
  \item \emph{L'Île aux fleurs} (Furtado, 1989)
 \end{itemize} &
 \begin{itemize}[leftmargin=*]
  \item \emph{After Work} (Pinget, 2020)
  \item \emph{Art TOTAL} (Excoffier, Pacotte, 2000)
  \item \emph{Contretemps} (Pollet, 1988)
  \item \emph{Le Temps des bouffons} (Falardeau, 1985)
 \end{itemize} &
 \begin{itemize}[leftmargin=*]
  \item \emph{My Father's Tools} (Condo, 2016)
 \end{itemize} \\
 \midrule
 \begin{description}[leftmargin=*]
   \item[\hspace{-2ex}] Éric
 \end{description} &
 \begin{itemize}[leftmargin=*]
  \item \emph{Nostalgie\dots{}} (Guzman, 2010)
 \end{itemize} &
 \begin{itemize}[leftmargin=*]
  \item \emph{L'Ordre} (Pollet 1973, 42 minutes)
  \item \emph{Le Temps des bouffons} (Falardeau, 1985)
  \item \emph{No Olvidar} (Agüero, 1982)
  \item \emph{Nuit debout} (Makengo, 2019)
 \end{itemize} &
 \begin{itemize}[leftmargin=*]
  \item \emph{À travers Jann} (Juge, 2019)
 \end{itemize} \\
 \midrule
 \begin{description}[leftmargin=*]
   \item[\hspace{-2ex}] Benjamin 
 \end{description}&
 \begin{itemize}[leftmargin=*]
  \item \emph{Nijuman no Borei\dots{}} (Périot, 2007)
 \end{itemize} &
 \begin{itemize}[leftmargin=*]
  \item \emph{El Cementerio\dots{}} (Yero, 2018)
  \item \emph{Ligne noire} (Olexa et Scalisi, 2017)
  \item \emph{Paraiso} (Guerrero, 2006)
  \item \emph{Petite Mort} (Bieber, 2016)
 \end{itemize} &
 \begin{itemize}[leftmargin=*]
  \item \emph{Urząd\dots{}} (Kieślowski, 1966)
 \end{itemize} \\
 \bottomrule
\end{xltabular}}

Le recueil des données a consisté à examiner, de manière systématique, l'ensemble des liens formés par ces recommandations personnalisées. Pour l'ensemble du panel, 44 films sont recommandés par l'algorithme auxquels s'ajoutent 11 films contrôles. La première partie du guide d'entretien cherche à expliciter les raisons qui ont conduit les abonnés à choisir un dispositif de réalisation plutôt qu'un autre parmi les 331 films qui composent le corpus. La deuxième partie se focalise sur les 5 recommandations proposées afin de faire émerger si ces liens entre les films sont perçus comme \emph{cohérents} ou \emph{incohérents}. Notre problématique vise à interroger la \emph{proximité, l'écart ou la distorsion} du dispositif documentaire choisi par l'abonné avec ceux des 5 films recommandés. L'évaluation se focalise ainsi sur le processus de \emph{construction de sens} en suivant l'axe de pertinence du dispositif. Il s'agit d'expliciter \emph{la demande faite au film} par l'abonné cinéphile, c'est-à-dire \emph{l'inférence} que celui-ci établit avec les différentes caractéristiques du documentaire. En résumé, l'approche compréhensive adoptée se focalise sur les \emph{logiques interprétatives} qui caractérisent les \emph{appropriations individuelles} de cette recommandation inédite.

\section{Réception collective de la recommandation}

\subsection{Une défiance de principe face aux algorithmes}

Le premier constat général que l'on peut dresser dans la réception de cette recommandation est qu'elle se heurte à un présupposé partagé par bon nombre d'abonnés de Tënk qui considère que toute recommandation algorithmique entraîne, \emph{de facto}, un enfermement dans une \enquote{bulle de filtre}. Dans ce positionnement volontiers critique vis-à-vis d'une recommandation automatisée, la disqualification de la démarche tend à s'affirmer au détriment de toute considération pour le résultat obtenu. Cette posture s'apparente souvent à un positionnement de principe contre ce qui est perçu comme une forme d'aliénation de la liberté humaine à une machine. C'est probablement oublier un peu vite que la recommandation humaine peine tout autant à faire consensus, qu'elle soit issue des critiques cinématographiques, ou qu'elle provienne de communautés d'amateurs via des sites collaboratifs ou les réseaux sociaux \citep{menard_systemes_2014}. Ce scepticisme sur les algorithmes fait également écho à une crainte, omniprésente ces dernières années dans l'espace public, sur l'usage mercantile des données personnelles. Or, ces critiques ciblent principalement des approches qui construisent des profils utilisateurs sur leurs traces d'usage contrairement à cette expérimentation qui s'appuie sur un autre paradigme. En effet, l'approche sémantique centrée sur l'œuvre cinématographique tente de décrire les spécificités du dispositif documentaire visionné par l'abonné sans exploiter son historique de consultation. Contrairement aux plateformes qui orientent la consultation vers une liste de lecture personnalisée jouée en flot continu, notre approche garantit une complète maîtrise à l'usager. Celui-ci reste anonyme et conserve l'entière liberté d'exploiter ou non les propositions qui lui sont faites. En dépit de cette loyauté vis à vis de l'abonné de Tënk qui s'inscrit dans les valeurs de la plateforme, le principe même de pouvoir guider l'abonné en fonction d'un dispositif plutôt qu'un autre rend septique une frange d'abonnés.

\begin{quotation}
 Le principe même de recommander, à partir de ce que j'ai vu, c'est une très mauvaise idée parce que le documentaire, par essence, c'est ouvrir des portes sur le monde. [\dots] À mon avis, un film n'est pas bon parce qu'il a un procédé cinématographique plutôt qu'un autre, un cadre formel. C'est comme si on disait que je voudrais un film où il y a beaucoup de jaune. Oui, c'est possible, il y aura beaucoup de jaune, mais ce n'est pas ça qui va le rendre bon. (A1, David, 58 ans, chercheur en mathématiques).
\end{quotation}

\subsection{Convergence entre l'indexation du dispositif et son interprétation par les abonnés}

Le deuxième constat général qu'il est possible d'établir est que les termes utilisés par les abonnés (en langage naturel) recoupent en grande partie les concepts formulés (en vocabulaire contrôlé) pour indexer le corpus de films. À titre d'exemple, Nadia définit le dispositif de \emph{Lift} (Isaacs, 2001) de la manière suivante:

\begin{quotation}
 La caméra est posée dans l'ascenseur avec le réalisateur derrière. J'ai un peu l'impression qu'il est là toute la journée, qu'il y a des gens qui appellent l'ascenseur, qui rentrent dedans. J'ai vraiment l'impression d'y être, d'assister au début d'une relation assez intime [\dots]. Le dispositif est limpide, les choix de mise en scène sont très assumés. Le fait que ce soit quelque chose de répétitif permet d'en être très conscient. (A6, Nadia, 28 ans, comédienne de théâtre).
\end{quotation}

Cette caractérisation du dispositif rejoint l'indexation où une série de 10 descripteurs ont été choisis comme le détaille le tableau suivant.

{\small
\begin{xltabular}[!ht]{\textwidth}{XXX}
 \caption{Corrélation entre l'indexation et l'interprétation du dispositif de \emph{Lift} (Isaacs, 2001).}\label{tab:lift}\\
 \toprule
 Descripteurs choisis lors de l'indexation par la documentaliste & Définition du descripteur dans le thésaurus (vocabulaire contrôlé) & Description du dispositif par l'abonné lors de l'entretien (langage naturel)\\
 \midrule
  \endfirsthead
 \toprule
 Descripteurs choisis lors de l'indexation par la documentaliste & Définition du descripteur dans le thésaurus (vocabulaire contrôlé) & Description du dispositif par l'abonné lors de l'entretien (langage naturel)\\
 \midrule
  \endhead 
1) Filmant en performance & La personne filmant met en scène ses propres actions dans le but de provoquer une situation. & \enquote{Il s'incruste à un endroit où il ne devrait pas être.}\\
\midrule
2) Huis clos & Le film est composé de plans tournés dans un seul et même lieu fermé. & \enquote{La caméra est posée dans l'ascenseur [\dots] toute la journée.}\\
\midrule
3) Cohérence formelle & Les plans qui composent le film sont liés par des similarités formelles (répétition d'un motif). & \enquote{C'est un dispositif qui me fait penser à la caméra cachée.}\\
\midrule
4) Protagoniste multiple & Le film compte plusieurs protagonistes qui ne forment pas un groupe en dehors du film. & \enquote{Il y a des gens qui appellent l'ascenseur [\dots] de classes sociales et de nationalités différentes.}\\
\midrule
5) Regard caméra du filmé & La personne filmée regarde la caméra et, à travers elle, la personne filmant autant que le public. & \enquote{Il y a aussi une triangulation avec la caméra parce qu'il y a des regards qui vont vers elle.}\\
\midrule
6) Filmant \emph{off} & La personne filmant manifeste sa présence physique hors du champ de la caméra sur le lieu et le temps du tournage. & \enquote{On entend le réalisateur leur parler.}\\
\midrule
7) Filmant complice & La personne filmant entretient une relation de complicité avec les personnes filmées. & \enquote{J'ai vraiment l'impression d'y être, d'assister au début d'une relation assez intime.}\\
\midrule
8) Dialogue filmant-filmé & La personne filmant et la personne filmée dialoguent ensemble. & \enquote{Il y a des interactions avec les habitants de l'immeuble.}\\
\midrule
9) Filmant opérateur & La personne filmant tourne elle-même des images du film. & \enquote{Le réalisateur est un interlocuteur direct des personnages.}\\
\midrule
10) Caméra portée & Un dispositif de prise de vue utilise la caméra comme un prolongement du corps. & \enquote{La caméra est posée [\dots] avec le réalisateur derrière}\\
\bottomrule
 \end{xltabular}
}

Dans cet exemple emblématique, l'abonnée relève chacun des aspects indexés, en insistant sur certains d'entre eux mais en passant plus rapidement sur d'autres. Le même type de correspondance pourrait être relevé par les autres membres du panel. Si des éléments de l'indexation sont parfois passés sous silence, chacun évoque au moins la moitié des traits saillants relevés lors de l'indexation.

Par ailleurs, la majorité du panel identifie que le film contrôle (qui ne possède aucun descripteur commun avec le film choisi) est très différent des autres documentaires recommandés comme le souligne par exemple David:

\begin{quotation}
 Ce film est définitivement le plus éloigné des autres. Je ne sais pas si vous l'avez fait exprès mais je me suis dit peut-être qu'ils ont pris un film appartenant à une tout autre famille pour voir si j'allais réagir? (A1, David, 58 ans, chercheur en mathématiques)
\end{quotation}

En bilan de cette première partie, nous constatons qu'il existe une convergence interprétative entre le travail humain d'indexation des dispositifs (effectuée en amont de la recommandation) et son interprétation par le panel d'abonnés cinéphiles (réalisée en aval de la recommandation). De plus, une lecture macroscopique des résultats indique qu'une majorité des recommandations proposées sont signifiantes pour les abonnés cinéphiles de Tënk. En effet, sur 44 films recommandés au panel, 28 liens sont jugés en cohérence avec le dispositif du film initial (soit 63\,\% des recommandations). Si ces résultats montrent que la recommandation fait globalement sens pour le panel, il semble désormais nécessaire de focaliser l'analyse à l'échelle individuelle. Ceci afin de s'intéresser aux différents modes d'appropriation spécifiques pour chaque abonné.

\section{Appropriation individuelle de la recommandation}

\subsection{Modes de lecture mobilisés par les abonnés}

En filiation avec l'approche sémio-pragmatique développée par Roger \citet{odin_espaces_2011}\footnote{Selon Roger \citeauthor{odin_espaces_2011}, \enquote{l'analyse sémio-pragmatique démonte l'illusion immanentiste qui laisse croire qu'il y a un seul texte et une seule lecture. [\dots] Toute lecture est un palimpseste de lectures} \citeyearpar[p.~140]{odin_espaces_2011}.}, l'analyse des entretiens met en exergue différents \enquote{modes de lecture} mobilisés par les abonnés cinéphiles pour justifier d'un goût pour une certaine démarche documentaire. À partir du paradigme du dispositif de réalisation présenté précédemment, les abonnés focalisent leur attention sur différentes compétences communicationnelles. L'expérimentation formelle, la relation du filmant à la personne filmée, la dimension performative de la mise en scène ou encore les qualités littéraires du commentaire, constituent autant de points de fixation possibles du regard de l'abonné. L'analyse suivante expose quatre déclinaisons de ces différentes postures interprétatives. Celles-ci ne sont pas exclusives les unes des autres et ne prétendent nullement à l'exhaustivité.

Premièrement, pour certains abonnés, le dispositif peut être interprété sur un \enquote{mode de lecture esthétique} \citep{odin_espaces_2011} qui s'incarne à travers des éléments visuels et sonores dotés de qualités plastiques et sensorielles. Les similarités formelles entre les films peuvent alors justifier, à leurs yeux, une cohérence de la recommandation. C'est le cas de Benjamin, particulièrement intéressé par les formes expérimentales du cinéma documentaire. Celui-ci choisit un film sur ce critère et attend des recommandations qu'elles s'inscrivent dans cette veine.

\begin{quotation}
    Ce qui me plaît dans le documentaire de création, c'est quand ça peut aller vers du film expérimental, dans une forme qui s'affranchit des limites imposées par la narration de fiction [\dots] Ce qui rapproche ces films, c'est leur liberté formelle. J'aime surtout cette forme d'animation qui n'est pas du tout illustrative. (A11, Benjamin, 26 ans, assistant-monteur)
\end{quotation}

Ce mode de lecture esthétique peut bien entendu s'hybrider avec d'autres critères d'appréciations. C'est le cas par exemple de Vincent qui déclare son admiration pour un court métrage qui associe créativité formelle et discours politique.

\begin{quotation}
    Ce film, c'est tout ce que j'aime: concision, ironie, efficacité, humour et intelligence au service d'une dénonciation en règle de l'absurdité et de l'inhumanité d'un système économique et social que nous subissons encore aujourd'hui. Le tout dans une forme extrêmement libre et pensée artistiquement. Il y a une vraie créativité. Ça ne ressemble à rien d'autre. C'est une espèce d'ovni au service de quelque chose de très important que je partage. (A9, Vincent, 62 ans, comédien et metteur en scène).
\end{quotation}

Deuxièmement, le mode de lecture adopté peut se cristalliser sur les interactions sociales entre la personne filmant et la personne filmée. Plusieurs abonnés portent ainsi une attention particulière à l'implication physique du cinéaste ou à la relation intime instaurée avec les personnes filmées. Vu sous ce prisme, les films qui conservent les marques d'une relation empathique entre le cinéaste et son sujet sont particulièrement appréciés.

\begin{quotation}
    Pour moi, filmer c'est accepter que l'autre laisse une trace en soi. Mais ce n'est pas souvent assumé et moi j'aime ce genre de posture-là [\dots]. À un moment, on voit carrément la réalisatrice qui se pose à côté du lit et qui vient taper l'épaule [du personnage principal]. Ces moments sont des pépites où ce n'est pas l'œil objectif de la caméra. (A2, Aline, 42 ans, éducatrice spécialisée).
\end{quotation}

Troisièmement, l'intérêt pour la place de la personne filmant peut conduire certains participants à apprécier des démarches cinématographiques où l'implication du cinéaste est visible à l'écran. Il s'agit généralement de films où le réalisateur fait partie intégrante de la diégèse et se met en scène dans le réel de manière performative.

\begin{quotation}
    Le réalisateur de \emph{Secteur 545} (Creton, 2004) se met dans une situation où il aide les exploitants qui ont des vaches à gérer la traite et la production laitière en Normandie. Tu vois l'entretien d'embauche dès le début, donc tu te dis qu'il a filmé dès son arrivée dans le métier. Après coup, tu comprends qu'il a fait pas mal de mise en scène, qu'il est également acteur. (A6, Nadia, 28 ans, comédienne de théâtre).
\end{quotation}

Quatrièmement, un autre critère d'appréciation permettant de manifester son intérêt pour le dispositif peut s'exprimer à travers la \emph{qualité littéraire du commentaire}. Adeline perçoit par exemple l'écriture de la voix \emph{off} du film \emph{Les Arbres qui marchent} (Colombo et De Luca, 2015) comme une \enquote{colonne vertébrale} qui relie le monde des arbres avec celui des hommes qui travaillent le bois. L'articulation d'une écriture littéraire avec des personnages qui manipulent de manière charnelle la matière première fait advenir une dimension spirituelle qui révèle \enquote{un esprit du bois}. Suivant le même critère d'appréciation, Éric voue une admiration pour l'écriture cinématographique de Patricio Guzman. Dans son documentaire \emph{Nostalgie de la lumière} (2010), le cinéaste parvient à réunir, dans un même espace désertique, l'observation du ciel et l'enquête des familles pour retrouver les corps des victimes disparues au Chili sous la dictature de Pinochet.

\begin{quotation}
    La voix \emph{off} de Guzman est très bien écrite. Elle permet de relier des choses très éloignées [\dots] Comme spectateur, cela nous amène à penser des questions existentielles sur notre statut d'êtres vivants dans le cosmos et sur le sens politique qu'on peut donner aux choses. Le film nous donne de l'espace pour avoir cette réflexion et nous laisse libre de divaguer sur ces questions. (A10, Éric, 44 ans, professeur-documentaliste).
\end{quotation}

\subsection{Intrication du dispositif de réalisation et de la thématique du film}

L'appétence pour la thématique abordée reste un élément déterminant dans le choix d'un film apprécié pour la qualité de son dispositif de réalisation. En effet, le sujet du film se trouve fréquemment intriqué avec le dispositif lorsque les abonnés jugent de la cohérence de la recommandation. Plusieurs abonnés révèlent ainsi une relation intime avec certains thèmes lorsqu'ils s'efforcent de décoder les différentes caractéristiques de la mise en scène adoptée par le réalisateur.

\begin{quotation}
    Le documentaire \emph{Dernières nouvelles du cosmos} (Bertuccelli, 2016) m'a vraiment marqué. C'est comme si, tout d'un coup, il avait touché un \emph{punctum}, un petit truc à l'intérieur de moi sur mon rapport à la folie. Mon père était éducateur spécialisé auprès de personnes handicapées. C'est des personnes que j'ai pu côtoyer et il y a quelque chose qui m'a toujours fasciné. Je me suis souvent posé la question de savoir si ce n'était pas nous les fous. (A2, Aline, 42 ans, éducatrice-spécialisée).
\end{quotation}

Il va de soi que le parcours biographique des abonnés, leur historicité, conditionne la relation intime qu'ils entretiennent, de manière plus ou moins consciente, à certaines thématiques. Alain déclare par exemple qu'il a adoré la série intitulée \emph{Le Village} (Simon, 2019) pour les convictions humanistes et politiques des protagonistes qui rejoignent son parcours professionnel dans le secteur médico-social ainsi que ses convictions en qualité d'ancien élu municipal à la culture d'une commune d'Île-de-France. Il apprécie particulièrement la manière dont le spectateur est placé en immersion, au plus près de l'action et des interactions sociales entre les différents personnages.

\begin{quotation}
    Le film se passe dans ce village où il y a une vie paysanne, une vie agricole, des récoltes, et il y a en même temps une autre activité [autour du cinéma documentaire] mais les deux ne se méprisent pas. [\dots] Je suis tellement en accord avec ce qui se dit qu'après, la forme disparaît un peu pour moi, elle devient invisible, ce qui est quand même une sacrée qualité. (A8, Alain, 83 ans, directeur-retraité de CMPP).
\end{quotation}

La frontière entre dispositif de réalisation et thématique est encore plus ténue lorsque les dispositifs des documentaires s'avèrent minimalistes. Il s'agit notamment de film qui relève du \enquote{cinéma direct} où \enquote{la forme est bien là, en creux, et elle est signifiante, mais l'affirmation du geste artistique y est moins saillante que dans les documentaires dits “à dispositifs”} \citep[p.~11]{zeau_cinema_2020}. La relative transparence du dispositif ne signifie pas pour autant l'absence de construction dans la représentation du réel mais Les situations observées font plus facilement ressortir la dimension thématique ou le sujet abordé par le cinéaste.
Influence contextuelle de l'interprétation individuelle

L'interprétation du dispositif se trouve également influencée très fréquemment par la manière dont la recommandation est perçue émotionnellement. Ainsi, les abonnés identifient une cohérence plus aisément lorsque le film proposé leur plaît particulièrement, et ceci quelles que soient les spécificités du dispositif. À l'inverse, un film qui enclenche une déception pousse le plus souvent les abonnés à décrire cette recommandation comme incohérente. C'est notamment le cas lorsque le film initial est choisi avec une grande exigence. Marianne s'avère par exemple particulièrement déstabilisée par certaines recommandations qui ne correspondent pas au niveau d'excellence qu'elle attend sur le plan esthétique.

\begin{quotation}
    Pour \emph{Fremd}, la qualité technique du tournage est vraiment médiocre. On n'a pas de plan magnifique. [\dots] Ça fait comme si quelqu'un était parti avec sa caméra avec un sujet qui lui tenait à cœur. Ce n'est pas un film qui me marque. (A5, Marianne, 68 ans, Institutrice-éducatrice retraitée).
\end{quotation}

Cette déception de l'abonnée peut être encore plus marquée si certaines caractéristiques du film recommandé vont jusqu'à rebuter l'abonnée, au point d'effacer, à ses yeux, toute autre forme de proximité possible entre les dispositifs. C'est ce que déclare par exemple Joséphine à propos de \emph{Libre} (Toesca, 2018):

\begin{quotation}
    Je trouve qu'on filme les migrants presque comme un troupeau, c'est-à-dire qu'on ne leur donne pas la parole. [\dots] Le réalisateur a un regard complètement condescendant. Ce n'est pas du tout le cas du film de Blaise Perrin, qui est à la même hauteur que ses personnages. Alors oui, on est dans le même thème, c'est deux personnes qui sont dans des situations de détresse et on filme quelqu'un qui les aide. Mais le lien s'arrête au sujet et en termes de forme, il n'y a pour moi aucun lien entre ces films. (A4, Joséphine, 33 ans, étudiante-programmatrice en cinéma).
\end{quotation}

Joséphine, mentionne également à plusieurs reprises ses grandes difficultés de concentration qui lui font quasi systématiquement abandonner le visionnage des films qu'elle entame, cela d'autant plus vite qu'elle détecte un indice sur un dispositif qui lui déplaît. Dès que ses attendus esthétiques en termes de dispositif sont déçus, elle devient hermétique à l'ensemble du documentaire, le considérant comme une recommandation incohérente.

\begin{quotation}
    Je regarde tout, dans le sens où je lance la lecture de tous les films. Et il y en a certains où au bout de 3 minutes, je vois que ça ne va pas du tout, ça ne me plaît pas. [\dots] Vraiment c'est très rare que je regarde un film sur Tënk jusqu'au bout. (A4, Joséphine, 33 ans, étudiante-programmatrice en cinéma).
\end{quotation}

En bilan de cette deuxième partie, il semble bien que chaque abonné mobilise une série de compétences communicationnelles pour apprécier le degré de similarité entre les différents dispositifs de réalisation proposés. Dans la continuité des travaux sémio-pragmatiques de Roger \citet{odin_espaces_2011}, ce processus de construction de sens s'appuie sur différents modes de lecture (esthétique, interactionnel, performatif, etc.). L'inférence faite au film par l'abonné se focalise ainsi sur l'une ou l'autre des dimensions qui constitue le dispositif de réalisation \citep[à paraître]{gantier_cartographier_2023}. Dit autrement, le spectateur cinéphile, en fonction de son historicité et d'un contexte donné\footnote{Selon le modèle sémio-pragmatique de Roger Odin, le contexte pour se définir comme \enquote{un faisceau de contraintes} qui conditionne différents modes de construction de sens se déployant selon différents espaces de communication \citep{odin_espaces_2011}.}, pose son regard, avec plus ou moins d'acuité, sur différents éléments du dispositif: la posture de la personne filmant, l'interaction entre filmant-filmé, l'intérêt du cinéaste pour la situation filmée, la manipulation voire la transformation des matériaux filmiques, les modes d'adresse au public, etc.

\section{Conclusion}

L'analyse de la réception d'une recommandation algorithmique (basée sur la similarité des dispositifs de réalisation) par des utilisateurs de la plateforme Tënk permet de dresser une série d'enseignements sur la médiation instrumentée du cinéma documentaire. De manière générale, une majorité des recommandations proposées dans le cadre de cette expérimentation fait sens pour une grande partie des abonnés cinéphiles. Ainsi, sur 44 documentaires recommandés, 28 sont jugés cohérents par le panel, soit 63\,\% des recommandations.

Quatre variables intriquées dans un processus info-communicationnel complexe sem\-blent conditionner l'appropriation de cette recommandation. Premièrement, la spécificité du dispositif documentaire choisi par l'abonné parmi les 331 films du corpus. Deuxièmement, le travail d'indexation qui détermine 10 descripteurs parmi 292 concepts proposés dans le thésaurus créé dans le cadre de cette étude. Troisièmement, la mise en calcul du jeu de donné qui établit une distance vectorielle entre les films. Quatrièmement, le processus interprétatif de l'abonné qui s'inscrit dans un faisceau de contraintes sociales et contextuelles au sein d'une communauté interprétative.

Si les résultats de cette preuve de concept sont globalement encourageants pour parvenir à guider le spectateur de films documentaires autrement que par les classifications thématiques existantes à ce jour, une série d'obstacles restent encore à franchir. Tout d'abord, l'appropriation se heurte au scepticisme que ce public témoigne vis-à-vis des algorithmes perçus comme une \enquote{boîte noire} qui nuirait, \emph{de facto}, à la diversité culturelle. Ensuite, il serait limitatif de vouloir faire complètement abstraction de la thématique abordée qui reste indubitablement présente dans l'appréciation du spectateur. Enfin, même adressée à un public d'initié, la recommandation semble souvent trop complexe à appréhender pour être perçue de manière immédiate. Le dispositif fait appel à une multitude de variables qui semblent plus ou moins perceptibles en fonction de l'échelle adoptée. Les entretiens soulignent ainsi le besoin d'expliciter cette recommandation, en amont par l'équipe éditoriale de la plateforme, pour être perçues et acceptés, en aval par le public.

En outre, cette expérimentation met en exergue les caractéristiques d'une pratique cinéphile bien particulière. Il s'agit, au sens où l'entend Laurent \citet{jullier_qu_2021}, d'une communauté qui use d'un paradigme commun pour produire un jugement normatif, c'est-à-dire qui s'accordent sur la hiérarchisation de critères qui déterminent la qualité d'un bon documentaire, sans pour autant nécessairement aimer les mêmes films. Ces critères participent à construire un \emph{paradigme cinéphile} spécifique au champ du documentaire de création. Celui-ci constitue \enquote{un mode d'accès au film, une façon particulière d'en extraire du sens [\dots] qui a pour conséquence de privilégier certains critères d'appréciation et d'en exclure d'autres. C'est un “certain regard”, un choix particulier d'attentes et de demandes qu'accompagne volontiers un vocabulaire adapté. La plupart des spectateurs en maîtrisent plusieurs, qu'ils utilisent implicitement en fonction du genre et du style du film, de leur humeur et de la situation} \citep[p.~207]{jullier_qu_2021}. De manière opératoire, le paradigme du dispositif de réalisation documentaire permet à cette communauté interprétative d'identifier les différents pôles qui constituent l'espace hétérogène du catalogue de Tënk. Sur le plan symbolique, il permet d'établir une série de marqueurs sémiotiques et de repères cognitifs pour positionner le film dans la variété des pratiques documentaires. En effet, chaque film diffusé sur Tënk développe un mode de représentation du réel singulier en fonction de la culture et de la pratique socio-professionnelle de son auteur (journalisme, arts contemporain, anthropologie, pédagogie, histoire, photographie, fiction, théâtre\dots{}) \citep{caillet_art_2017}.

En termes de perspectives de recherche, d'autres scénarios d'usage que celui des abonnés cinéphiles de Tënk pourraient être imaginés. Cette approche centrée sur l'indexation du dispositif des œuvres documentaires pourrait par exemple s'adresser à des programmateurs de festival pour les aider à élaborer une programmation originale et non thématique, ou encore à des formateurs qui encadrent des résidences d'écriture afin d'accompagner des auteurs-réalisateurs à développer un projet de film. Dès lors, l'enjeu ne serait plus de proposer une recommandation qui irait de soi, mais davantage de permettre à ce public d'initié de naviguer, de manière supervisée, dans de vastes catalogues de plusieurs milliers de titres catégorisés selon une description détaillée de leurs dispositifs de réalisation.

\renewcommand{\refname}{Bibliographie}
\bibliographystyle{apacite}
\bibliography{AlgoDoc-bibtex}

\end{document}